\documentstyle[12pt]{article}
\newcommand{\sect}[1]{\section{#1}\setcounter{equation}{0}}

\def\apm{\alpha'}
\begin{document}
\bigskip
\hskip 3.7in\vbox{\baselineskip12pt
\hbox{NSF-ITP-97-027}\hbox{hep-th/9704029}}
\bigskip\bigskip\bigskip\bigskip

\centerline{\large \bf Membrane Scattering with M-Momentum Transfer}

\bigskip\bigskip
\bigskip\bigskip

\centerline{\bf Joseph Polchinski} 
\medskip
\centerline{Institute for Theoretical Physics}
\centerline{University of California}
\centerline{Santa Barbara, CA\ \ 93106-4030}
\centerline{e-mail: joep@itp.ucsb.edu}
\bigskip
\centerline{\bf Philippe Pouliot} 
\medskip
\centerline{Department of Physics}
\centerline{University of California}
\centerline{Santa Barbara, CA 93106}
\centerline{e-mail: pouliot@sarek.physics.ucsb.edu}

\begin{abstract}
\baselineskip=16pt
Membrane scattering in m(atrix) theory is related to dynamics in
three-dimensional $SU(2)$ gauge theory, with transfer of $p^{11}$ being an
instanton process.  We calculate the instanton amplitude and find precise
agreement with the amplitude in eleven dimensional supergravity.
\end{abstract}
\newpage
\baselineskip=18pt

\sect{Introduction}

Matrix theory is a notable proposal for the fundamental degrees of freedom
and their Hamiltonian~\cite{bfss}.  Roughly speaking, it is proposed that of
those structures known in the weakly coupled string limits, it is the
Dirichlet zero-branes of the IIA string~\cite{dbranes} and their low-energy
Hamiltonian~\cite{witbound,dham} that actually have a much wider range of
validity.  More specifically, for states of large M-momentum (momentum in
the eleven-direction of M-theory) these are supposed to give a complete
description, even in the limit of large string coupling where
eleven-dimensional Lorentz invariance reappears.

The strongest evidence that this is a step in the right direction is the
connection with the supermembrane of eleven-dimensional
supergravity~\cite{smem,memham,town}.  Among the other tests
are comparisons of various matrix theory scattering amplitudes with those
of eleven-dimensional supergravity~\cite{bfss},\cite{aharberk}-\cite{lif5},
but only at vanishing M-momentum transfer.  Processes with transfer of
M-momentum are qualitatively different and seemingly much harder to
calculate.  In the matrix theory description, the former involve only a
loop of virtual open string, while the latter require transfer of
zero-branes and so depend on the still-mysterious supersymmetric bound
states of zero-branes.  At this point it is possible that matrix theory is
some mutilation of the correct theory, which reproduces the amplitudes of
vanishing M-momentum transfer but not more general amplitudes.  Of course
to single out one momentum mode in this way is highly nonlocal, but this
is just the point: in matrix theory, locality is not at all manifest,
especially in the M-direction.

In this paper we report some progress in this direction.  We are able to
calculate the amplitude for scattering of transverse supermembranes with
exchange of one unit of M-momentum.  The result is in detailed agreement
with the amplitude in eleven-dimensional supergravity.
The calculation is
valid in a limit where the impact parameter and the string coupling are
taken large together.  As with other scattering amplitudes, to reach the true
M-theory limit of large string coupling at fixed impact parameter will
require a supersymmetric nonrenormalization theorem.

The reason that we are able to make progress is that it appears to be
simpler to boost a membrane in the M-direction than a
graviton~\cite{aharberk,lifmat}.  The latter requires increasing the
number of zero-branes in the bound state, and no simple scaling properties
are yet known.  The former merely increases an internal magnetic field and
the Hamiltonian scales in a simple way.  The scattering calculation then
reduces to an instanton calculation in
$(2+1)$-dimensional gauge theory.

In section~2 we obtain the matrix theory action for the two-membrane system. 
In section~3 we relate the matrix theory amplitude to an instanton
calculation, and find detailed agreement with the eleven-dimensional result.
Section~4 contains a brief discussion.

\sect{The Membrane Action}

We consider a pair of membranes infinitely extended in the $2,3$-directions,
with some separation and relative motion in the $4, \ldots,
10$-directions and both moving with large velocity in the 11-direction. 
Minkowski time is
$x^0$ and Euclidean time $x^1 = - i x^0$.  In the IIA string theory,
the membrane degrees of freedom are described by the Dirac-Born-Infeld
lagrangian~\cite{DBI}
\begin{eqnarray}
S &=& -\tau_2 \int d^{3}x\, {\rm Tr}\det\!^{1/2}\left(-\eta_{\mu\nu} -
\partial_\mu X^i \partial_\nu X^i + 2\pi\apm F_{\mu\nu} \right) 
\nonumber\\
&&\qquad + O([X^i,X^j]^2) + {\rm fermions}\ .
\label{dact}
\end{eqnarray}
The precise Born-Infeld form of the commutator and fermionic terms will not
be needed.  Here $\mu,\nu=0,2,3$ and $i = 4,\ldots,10$.  The membrane tension
is~\cite{dbranes}
\begin{equation}
\tau_2 = \frac{1}{4\pi^2 \apm^{3/2} g}
\end{equation}
with $g$ the string coupling.  The transverse coordinates $X^i(x^\mu)$ and
the field strength $F_{\mu\nu}$ are $2\times 2$
matrices~\cite{witbound}.  

When the
membranes are separated, with $[X^i,X^j] = 0$, then up to a gauge
transformation
\begin{equation}
X^i= \left[ \begin{array}{cc} x_1^i & 0 \\ 0 & x_2^i \end{array} \right]
\end{equation}
breaks $U(2)$ to $U(1)\times U(1)$.
The embedding
coordinate for the eleventh dimension comes from the scalar dual
to a gauge field in $2+1$ dimensions~\cite{eleven,town}. 
Restricting to the unbroken $U(1) \times U(1)$, treat the field strength
rather than vector potential as independent and add to the action a term
\begin{equation}
S'= \frac{1}{2}
\int d^{3}x\, \epsilon^{\mu\nu\rho}\lambda_r \partial_\mu F_{r\nu\rho}
\end{equation}
to enforce the Bianchi identity, with $r$ indexing the two membranes. 
Integrating out
$F_{r\nu\rho}$ gives the Lagrange multiplier fields $\lambda_r$ kinetic
terms. From the normalization of the kinetic term one deduces that $\lambda =
2\pi\apm \tau_2 X^{11} = X^{11}/2\pi\apm^{1/2}g$ for each membrane.  The
periodicity of each $\lambda$ is simply $\lambda \sim \lambda + 1$.  This
follows because the Dirac quantization condition requires the total $U(1)$
flux on any membrane to be conserved mod $2\pi$, and as a
check the periodicity $X^{11} \sim X^{11} + 2\pi\apm^{1/2}g$ gives the
correct relation between the zero-brane mass $\tau_0 = 1/\apm^{1/2}g$ and the
Kaluza-Klein spectrum.

We will take the membranes to have equal velocity in the eleven-direction.
This corresponds to a field strength 
\begin{equation}
F_{23} = \frac{I f}{2\pi\apm}  \label{abelf}
\end{equation}
with $I$ the $2\times 2$ identity and $f$ a constant.  From the field
equation for $S + S'$ one then has
\begin{equation}
\dot X^{11} = f/\sqrt{1+f^2}\ .
\end{equation}
As a check, this gives a momentum density $\Pi_{11} = \tau_2 f$, implying
a zero-brane charge density $\tau_2 f/\tau_0$.  This is in agreement with
the coupling of the D-brane to the Ramond-Ramond field~\cite{rrcs},
proportional to
$\tau_2(C_{(3)} + 2\pi\apm F \wedge C_{(1)})$ for the two-brane and
$\tau_0 C_{(1)}$ for each zero-brane.

The above D-brane description of the membranes is superficially
different from the matrix theory description as a non-Abelian state of
zero-branes~\cite{bfss,memham}, but the equivalence should follow as in
ref.~\cite{bss,lifmat}.  We thus take
the above as our tentative definition of matrix theory for this system. 
That is, while the above description is known to be valid only for weakly
coupled strings, we conjecture that for  large boosts in the M-direction ($f
\gg 1$), it remains a valid description even for large string coupling where
eleven-dimensional Lorentz invariance should reappear.  We then compare
scattering amplitudes with those of eleven-dimensional supergravity.

We will be studying nearly supersymmetric states, small deviations from the
parallel motion~(\ref{abelf}).  Naively one might expect the $SU(2)$ degrees of
freedom to decouple from the $U(1)$ background, since they are neutral
under the center-of-mass $U(1)$.   However, the endpoints of each open
string carry equal and opposite charges.  For a strong magnetic
field, where the number of flux quanta per string area is large, one would
expect the dynamics to be substantially affected by the field.  Indeed,
these complications are precisely taken into account by the DBI
action~\cite{callan}, so we need merely do the obvious thing, to
expand the DBI action~(\ref{dact}) around the
background~(\ref{abelf}).  The result is
\begin{equation}
S = -\tau_2 \gamma^{-1} \int d^{3}x\,
\sqrt{- \det \hat\eta}\, {\rm Tr}\left\{ \frac{1}{2} \partial_\mu
X^i \partial^{\hat\mu} X^i + \pi^2\apm^2 F_{\mu\nu}F^{\hat\mu\hat\nu}
\right\}\ .
\label{redact}
\end{equation}
We have defined the Lorentz boost factor
\begin{equation}
\gamma\ =\ \left( 1 - v_{11}^2 \right)^{-1/2}
\ =\ (1 + f^2)^{1/2}
\end{equation}
and the metric
\begin{equation}
\hat\eta_{\mu\nu} = 
\left[ \begin{array}{ccc} -1 & 0 & 0 \\
0 & \gamma^2 & 0 \\ 0 & 0 & \gamma^2
\end{array} \right]\ .
\end{equation}
The hatted metric has been used to raise and lower indices, as indicated.

Introduce the new coordinate $y^m$,
\begin{equation}
y^0 = x^0, \quad y^{2,3} = \gamma x^{2,3}
\end{equation}
in terms of which the metric is just $\eta_{mn}$.  Define $X^i = 2\pi\apm
\phi^i$, and convert from $SU(2)$ matrix notation to vector notation,
$\phi^i = \frac{1}{2} \sigma^a \phi^{ai}$.  The action becomes
\begin{equation}
S = -\frac{\apm^{1/2}}{2g\gamma} \int d^{3}x\,
\left\{ \frac{1}{2} \partial_m
\phi^{ai} \partial^m \phi^{ai} + \frac{1}{4} F^a_{mn}F^{amn}
\right\}\ ,
\end{equation}
an ordinary $SU(2)$ Yang-Mills-Higgs action with coupling $e^2 =
2g\gamma/\apm^{1/2}$.

Thus far we have used the natural string parameters $\apm$ and $g$.  The
relation to M theory parameters is
\begin{equation}
R_{11} = \apm^{1/2} g\ ,\qquad M_{11}^{-3} = \apm^{3/2} g\ .
\end{equation}
Here $R_{11}$ is the radius of the M-direction, and $M_{11}$ is the
eleven-dimensional Planck scale up to a numerical factor.

\sect{Membrane Scattering}

For scattering with impact parameter $X^i \sim b$, the effective
dimensionless coupling is
\begin{equation}
\frac{e^2}{\phi} \sim  \frac{\gamma g\apm^{1/2}}{b} \sim \frac{\gamma
R_{11}}{b}\ .
\end{equation}
This is large in the M theory limit of large $R_{11}$ with fixed $b$, so
the effective $SU(2)$ Yang-Mills theory is strongly coupled.  Recently,
beginning with refs.~\cite{seiwit}, many exact results have been derived
for three-dimensional gauge theories.  We began this work with the hope
that these exact results might be used directly to test matrix theory
predictions for membrane scattering.  However, the exact results are
primarily for $v^2$ terms in $d=3$, $N=4$ supersymmetry, whereas the
two-membrane system has $N=8$ supersymmetry where the leading terms are
$v^4$.  As with other tests of matrix theory
scattering~\cite{bfss},\cite{aharberk}-\cite{lif5}, we must wait for a
more complete understanding of the constraints that supersymmetry places
on these
$v^4$ terms. 

One might hope to study matrix theory questions with less supersymmetry. 
The simplest way to reduce the supersymmetry (and specifically to remove
the adjoint hypermultiplet) is to consider membranes trapped at the fixed
point of a K3 orbifold.  However, this changes the M theory picture
substantially.  Blowing up the fixed point slightly, a trapped D
two-brane is evidently a D four-brane wrapped on the collapsed two-sphere.
In M theory this becomes a five-brane, extended in the M direction.  So
the membrane is no longer a localized probe in this direction.  This
defeats our present purpose; it may be interesting to return to this
point later.

What we shall do is to consider instead the regime $b \gg R_{11}\gamma$
where reliable calculations can be made.  As with other matrix
theory scattering calculations, we will have to assume that
nonrenormalization theorems allow the result to be extended to the M
theory limit.  However, we will already be
able to see local eleven-dimensional physics in the region of validity
of the calculation.

\subsection{Instanton Calculation}

The relation between magnetic flux on the membrane and M-momentum means
that in a scattering with one unit of M-momentum transfer, the change in
the flux is
\begin{equation}
\int dx^2 dx^3 \left[F_{23}(x^0 = \infty) - 
F_{23}(x^0 = -\infty) \right] = 2\pi \sigma^3\ .
\end{equation}
That is, the integral over the sphere at infinity of the flux
in the unbroken $U(1) \subset SU(2)$ is non-zero.  This is therefore an
$SU(2)$ instanton process, since the instanton of three-dimensional gauge
theory is the same as the magnetic monopole in three spatial
dimensions~\cite{moninst}.  Instanton effects were considered in
$d=3$ $N=1$ and $N=2$ supersymmetry in ref.~\cite{AHW}.  Just as this is
being written, an explicit calculation for $N=4$ supersymmetry has
appeared~\cite{dorey}, which has substantial overlap with the calculation
below.

The Euclidean $N=8$ Yang-Mills action, now including the commutator and
fermionic terms, can be conveniently written as the dimensional reduction
of the $d=10$ theory,
\begin{equation}
S = \frac{1}{e^2} \int d^3 y \left\{
\frac{1}{4} F_{MN} F_{MN} + \frac{i}{2} \bar\psi \gamma_M D_M \psi 
\right\}\ .
\end{equation}
Here $M,N$ run over $m = 1,2,3$ and $i = 4, \ldots, 10$,
with $A_M \equiv (A_m, \phi_i)$.  The $SU(2)$ vector index $a$ is
suppressed.  The gamma matrices can be taken as
\begin{equation}
\gamma_m = \sigma_m \otimes 1 \otimes \tau^1, \qquad \gamma_i = 1
\otimes \Gamma_i \otimes \tau^2\ .
\end{equation}
Here $\sigma_m$ and $\tau^{1,2}$ are both used to denote the Pauli
matrices, but in the first and third factors respectively,\footnote
{Earlier $\sigma^a$ was also used for the gauge $SU(2)$.} while
$\Gamma_4,\ldots,\Gamma_{10}$ are the (pseudo-Majorana) $8\times 8$
$SO(7)$ gamma matrices.  The Weyl condition is
$\psi = \tau^3 \psi$, while the Majorana property
means that $\bar \psi = \psi^T \gamma_2$.

The supersymmetry variation of $\psi$ is
\begin{eqnarray}
\delta\psi^a &=& \frac{i}{2} F^a_{MN} \gamma_M \gamma_N \epsilon\\
&=& - (B^a_m + D_m \phi^a_i \Gamma_i) \sigma_m \epsilon
\label{susyvar}
\end{eqnarray}
with $\tau^3 \epsilon = \epsilon$.  Here $B^a_m = \frac{1}{2}
\epsilon_{mnp}F_{np}$.  To be specific let us consider first membranes
separated in the 4-direction with no transverse velocity, so that $\phi_4$
breaks the gauge $SU(2)$ to $U(1)$ and the transverse $SO(7)$ to $SO(6)$. 
The instanton lies in the same $SO(7)$ direction, $\phi_4^a$.  It follows
from the supersymmetry variation~(\ref{susyvar}) that for an instanton
satisfying the BPS condition~\cite{ps}
\begin{equation}
B^a_m = D_m \phi^a_4 \label{BPS}
\end{equation}
the eight supersymmetries with $\Gamma_4 \epsilon = - \epsilon$ are
unbroken.  The other eight supersymmetries give fermionic zero modes,
\begin{equation}
\psi^a = B^a_m \sigma_m \epsilon \label{fzm}
\end{equation}
for $\Gamma_4 \epsilon = + \epsilon$.
The instanton action is $S_{\rm cl} = s_0/e^2$ where
\begin{equation}
s_0 = \frac{1}{2}\int d^3y\,\left\{ B^2 +(D\phi)^2 \right\}
= \int d^3y\,D_m (B_m^a \phi^a_4)\ =\ 4\pi \phi_0\ ,
\end{equation}
with $\phi_0$ the asymptotic value of $(\phi_4^a \phi_4^a)^{1/2}$.

Expand around the background, $A_M = A_{M\rm cl} + a_M$, and add to the
Lagrangian the background gauge-fixing term $(D_M a_M)^2/2e^2$.  The
action becomes
\begin{equation}
S = \frac{s_0}{e^2} + \frac{1}{2e^2} \int d^3 y
\left\{ a_M \Delta_{MN} a_N + i \bar\psi \gamma_M D_M \psi
- b D^2 c \right\} + O(a^3)
\end{equation}
with $b$ and $c$ the Fadeev-Popov ghosts and
\begin{equation}
\Delta_{MN} a_N = - D^2 a_M + 2 F_{MN} a_N\ .
\end{equation}
The square of the Dirac operator is
\begin{equation}
(i \gamma_M D_M)^2 = -D^2 + i (1 + \Gamma_4) B_m \sigma_m\ .
\end{equation}
That is, this sixteen-component operator reduces to eight copies of the
one-component operator $-D^2$ and four copies of the two-component
operator
\begin{equation}
\Delta = -D^2 + 2 i B_m \sigma_m\ .
\end{equation}
Also, if one assembles the bosonic fluctuations $a_{1,\ldots,4}$ into a
matrix
$M(a_M) = a_4 - i a_m \sigma_m$, one finds that
\begin{equation}
M(\Delta_{MN} a_N) = \Delta M(a_M)\ ,
\end{equation}
so that the bosonic fields give two copies of $\Delta$, from the the
columns of $M$.  The components $a_{5,\ldots,10}$ give six copies of
$-D^2$.  Thus all determinants involve only two differential
operators~\cite{dadda}, namely
$-D^2$ (with no zero modes) and
$\Delta$ (with two zero modes~\cite{callias}).  Defining the measure so
that the gaussian path integral of
\begin{equation}
\frac{1}{2e^2} \int d^3 y
\left\{ \phi_0^{2} a_M  a_M + \phi_0 \bar\psi \psi
+ \phi_0^{2} b c \right\}
\end{equation}
is unity, the nonzero mode determinants are
\begin{equation}
(\det -\phi_0^{-2} D^2)^{-3 + 2 + 1} (\det{}' \phi_0^{-2} \Delta)^{-1 + 1 +
0} = 1\ ,
\end{equation}
the exponents coming respectively from the bosons, fermions, and
ghosts.\footnote
{Of course the final amplitude cannot depend on the definition of the
measure, but to see this in detail depends on a subtle non-cancellation
of the nonzero modes.  Happily, the authors of ref.~\cite{dorey} have
spared us from explaining this.}

According to the above discussion, there will be four bosonic zero modes.
Three are from translations,
\begin{equation}
a_4^{(m)} = D_m \phi_{\rm cl},\qquad a_n^{(m)} = F_{mn\,\rm cl}\ .
\end{equation}
The fourth is a rotation in the unbroken $U(1)$ with gauge parameter
$\lambda^a = \phi_4^a/\phi_0$,
\begin{equation}
a_4^{(0)} = 0,\qquad a_n^{(0)} = \phi_0^{-1} D_n \phi_{\rm cl}\ .
\end{equation}
This is not a gauge rotation, being nontrivial at infinity, and so gives
rise to a normalizable zero mode.  The finite transformation is trivial
at infinity for $\alpha = 2\pi$.
Using the BPS condition~(\ref{BPS}), one finds that
\begin{eqnarray}
\int d^3y\, a_M^{(m)} a_M^{(n)} &=& s_0 \delta_{nm}
\nonumber\\
\int d^3y\, a_M^{(0)} a_M^{(0)} &=& \phi_0^{-2} s_0\ .
\end{eqnarray}
The gaussian normalization thus determines the bosonic zero mode measure
to be
\begin{equation}
\frac{\phi_0^3 s_0^2}{4\pi^2 e^4}\, d^3 y\, d\alpha
\to \frac{8\pi \phi_0^5}{ e^4}\, d^3 y\ ,
\end{equation}
where the integral over $\alpha$ just gives its range $2\pi$.

Similarly the eight fermionic zero modes~(\ref{fzm}) are normalized
\begin{equation}
\frac{e^8}{\phi_0^4 s_0^4} \prod_{I=1}^8 d\theta_I\ . \label{fzmnorm}
\end{equation}
The simplest nonzero amplitude, with eight massless fermions,
is then
\begin{equation}
\left\langle \prod_{s=1}^8 \bar\zeta_s\tilde\psi(y_s)
\right\rangle = \frac{e^4 e^{-4\pi \phi_0/e^2}}{32\pi^3 \phi_0^3}
\int d^3 y\, \det K_{sI}(y)
\end{equation}
where $K_{sI}(y) = \bar\zeta_s \sigma_m
\epsilon_I \tilde B_m(y_s - y)$.  The $\epsilon_I$ run over the eight
spinors with $\Gamma_4 = +1$.  We use a tilde to denote contraction
with the normalize Higgs field, e. g.  $\tilde\psi
= \phi_{\rm cl}^a \psi^a /\phi_0$.

To express the result as an effective operator, note that the fermionic
propagator is $e^2 y_m\sigma_m/4\pi y^3$ while $\tilde B_m = y_m/y^3$.
The effective operator is thus
\begin{equation}
\frac{2^{11} \pi^5}{e^{12}} \int d^3 y\,
\phi^{-3} e^{-4\pi \phi/e^2 + i \lambda}
\prod_{\beta=1}^8 \psi_\beta\ .
\end{equation}
The product runs over the eight spinor components with $\Gamma_4 = +1$.
We have included the dependence on the dual scalar
$\lambda$~\cite{moninst,AHW}.

To compare with the gravitational result, it is more convenient to
consider the operator with no fermions but four powers of the membrane
velocity.  This is related to the above by supersymmetry, but we will
obtain it directly from the instanton calculation.  The instanton solution
depends on the asymptotic values $\tilde\phi_i$ of the moduli.  Let
$\varphi_{\rm cl}(y,y',\tilde\phi)$ refer to the classical value of a
generic field
$\varphi(y)$ in the instanton solution that is centered at
$y'$ with asymptotic moduli $\tilde\phi$.  We expand around the
quasisolution
\begin{equation}
\varphi_{\rm cl}(y,y',\tilde\phi(y^1))  
\end{equation}
where $\tilde\phi_i(y^1) = b_i + u_i y^1$.  Thus $b_i$ is the impact
parameter and $u_i$ the Euclidean velocity, which we take to be small.
Away from the instanton this reduces to linear motion of the membranes. 
We can choose a frame in which $b_i$ is in the 4-direction and $u_i$ in the
5-direction.  

To saturate the fermionic path integral we need four powers
of the action
\begin{equation}
\delta S = \frac{i}{e^2} \int d^3 y\, \bar\psi \gamma_M \delta D_M \psi
\end{equation}
where the velocity enters into the action through the change in the Dirac
operator, $\delta D_M$.  The relevant overlap integrals are then
\begin{equation}
\frac{i  u_i}{e^2} \int d^3 y\, y^1 \bar\psi_{I} \gamma_M 
\frac{\delta D_M}{\delta \tilde\phi_i} \psi_{J} \label{overlap}
\end{equation}
where again the fermionic zero modes are $\psi^a_{I} = B^a_m \sigma_m
\epsilon_I$.  Since $\gamma_M D_M \psi_{J} = 0$ we have also 
\begin{equation}
\frac{\delta }{\delta \tilde\phi_i} (\gamma_M D_M \psi_{J}) = 0\ .
\end{equation}
We can therefore move the modular derivative over to $\psi_{J}$ and then
integrate by parts to turn the overlap integral into
\begin{equation}
\frac{i u_i}{e^2} \int d^3 y\,  \bar\psi_{I} \gamma_1 
\frac{\delta }{\delta \tilde\phi_i} \psi_{J}\ . \label{over2}
\end{equation}
Now let us note that for spinors $\epsilon$ and $\epsilon'$ having
$\Gamma_4 = +1$, the overlap $\bar\epsilon \gamma_1 \epsilon'$ vanishes.
The nonzero contribution to the overlap~(\ref{over2}) must come from the
rotation of $\epsilon_J$.  The rate of angular rotation in the 45-plane
is $u_{\perp}/\phi$, so the overlap is this, times a $\frac{1}{2}$ from
the spinor rotation matrix, times $s_0$ from the spatial integral as
before.  Each spinor has a nonzero overlap with the rotation of exactly
one other, giving the overlap to the fourth power.  Rejoining the earlier
calculation after eq.~(\ref{fzmnorm}), we have the final amplitude
\begin{equation}
\frac{\pi e^4  }{2 }
\int d^3 y\, \frac{ (u_{\perp}^2)^2 }{ \phi^3} e^{-4\pi \phi/e^2 + i
\lambda} \ .
\end{equation}

To conclude this section we use the results of section~2 to express the
amplitude in terms of M theory quantities,
\begin{equation}
\frac{1 }{16 R^3_{11} M_{11}^3 }
\int d^3 x\, \frac{ (\dot X_{\perp}^2)^2 }{ X^3} e^{-(X/\gamma - i
X^{11})/R_{11}}  \label{instscat}
\end{equation}
where $X$ is the transverse separation in ten dimensions.

\subsection{Supergravity Calculation}

The matrix theory amplitude is to be compared with the scattering in low
energy supergravity.  One can consider the action for one membrane moving
in the field of another.  Because $R_{11}$ remains finite, we must
include also the fields of the images.  The `source' membrane and its
images are at
\begin{equation}
x^{11} = v_{11} x^0 + 2\pi n R_{11}\ .
\end{equation}
We find it convenient to boost to the rest-frame in the M-direction,
\begin{eqnarray}
x^{11 \prime} &=& \gamma(x^{11} - v_{11} x^0)\ =\  2\pi n \gamma R_{11}
\nonumber\\
x^{0 \prime} &=& \gamma(x^{0} - v_{11} x^{11})\ =\ 
\gamma^{-1} x^{0} - 2\pi n \gamma R_{11} v_{11}.
\end{eqnarray}
In this frame only the transverse velocity of the `test' membrane remains.
The action for the test membrane is
\begin{equation}
-\tau_2 \int d^3x' \det{}^{1/2} (h_{\mu\nu}) + i \mu_2 \int H
\end{equation}
with the induced metric 
\begin{equation}
h_{\mu\nu} = g_{\mu\nu} + \partial_\mu x^i
\partial_\nu x^j g_{ij}\ .
\end{equation}
The metric of the source membrane is~\cite{duffstelle}
\begin{eqnarray}
&&g_{\mu\nu} = f^{-2/3}(r) \eta_{\mu\nu}, \qquad
g_{ij} = f^{1/3}(r) \delta_{ij} \nonumber\\
&&f(r) = 1 + \frac{r_0^6}{r^6}\ ,
\end{eqnarray}
with $r$ the transverse separation.

Expanding in to fourth order in the velocities $v^i = \partial_0 x^i$,
\begin{equation}
\det{}^{1/2} (h_{\mu\nu}) = f(r)^{-1} - \frac{1}{2} v^2 
- \frac{1}{8} f(r) (v^2)^2\ . \label{velex}
\end{equation}
The velocity-independent term cancels the antisymmetric tensor
interaction.  The $v^2$ term is position-independent, so the leading
interaction is the $v^4$ term
\begin{equation}
\frac{\tau_2 r_0^6}{8 r^6} (v^2)^2\ . 
\end{equation}
In order to determine the value of $\tau_2 r_0^6$ we compare with the
gravitational contribution to the static force between D
two-branes in the IIA string~\cite{dbranes}, 
$3 \apm/2 X^5$.
To compare to the velocity-independent term $\tau_2 r_0^6 / r^6$ in
eleven dimensions, we note that at long distance the
interaction in the IIA string comes from the smeared sum over the periodic
images,
\begin{equation}
\frac{3\apm}{2 X^5} = \int \frac{dx^{11}}{2\pi R_{11}} \frac{\tau_2
r_0^6}{(X^2 + R_{11}^2)^3}
\end{equation}
or $\tau_2 r_0^6 = 8\apm R_{11} = 8M_{11}^{-3}$.

The $v^4$ interaction is then
\begin{equation}
\frac{1}{M_{11}^3} \int d^3x'\, (\partial_0' X^i \partial_0' X^i)^2
\sum_{n = -\infty}^\infty
\left( X^2 + [2\pi n R_{11} - X^{11}]^2 \gamma^2 \right)^{-3}\ .
\end{equation}
To pick out the term with one unit of M-momentum, multiply by
$e^{-i X^{11}/R_{11}}$ and average from $0$ to $2\pi R_{11}$.  The sum on
$n$ can be used to extend the integral from $-\infty$ to $\infty$,
with the leading result at large $r$ being
\begin{equation}
\frac{1}{16 \gamma^3 R_{11}^3 M_{11}^3} \int d^3x'\, \frac{(\partial_0'
X^i
\partial_0' X^i)^2}{X^3} e^{-(X/\gamma - iX^{11})/R_{11}}\ .
\end{equation}
Boosting back to the lab frame introduces a factor of $\gamma^3$, giving
precise agreement with the matrix theory
result~(\ref{instscat}).\footnote{The instanton amplitude involves only
the transverse velocity, whereas the supergravity calculation involves
the total velocity.  However, terms with the radial velocity can be
converted by parts into second-derivative terms, to which the instanton
calculation is insensitive.}
In
particular, the exponential suppression of the instanton result has a
simple spacetime origin.  At long distance the fields of the periodic
images overlap, and so the $x^{11}$ dependence falls exponentially.

\sect{Discussion}

We have tested the eleven-dimensional Lorentz invariance of
matrix theory beyond the known results on the classical Lorentz
invariance of the supermembrane action and on scattering at zero
M-momentum transfer. 

It is important to determine the extent to
which the result is restricted by supersymmetry.  It is clear that the
instanton calculation is unchanged if additional massive degrees of
freedom are added to the theory, so these cannot be excluded.  It may be
that supersymmetry determines fully the form of the amplitude, given its
$X^{11}$-dependence and the perturbative $v^4$ term.  Supersymmetry alone
however cannot determine the absolute normalization, since it linearizes
to this order.  However, supersymmetry plus a nonsingularity condition
(which would appear at $X \to 0$, deep in the M theory regime) might do
so.  The $N=4$ case~\cite{dorey} is a clear illustration of this.

It may be useful to study further the physics in the regime of validity of
the present calculation, $X \gg R_{11}\gamma$.  Although the full
Lorentz invariance is not visible there, any complete theory must
include this regime, and at least some matrix-theory processes are weakly
coupled.  Thus one should be able to consider less supersymmetric
processes, e. g.~\cite{ballar}.

It remains to be seen whether this result for membranes gives any
insight into the corresponding graviton scattering amplitudes.  It is
interesting to note that whereas the zero-brane system is strongly
interacting, the fluctuations of a single membrane are weakly interacting
in the infrared.  It may be that (compact) membrane-like states are in some
sense attractive in the large-$N$ limit of the zero-brane system, so that
the zero-brane bound state becomes in some sense `membrane-dominated.'

\subsection*{Acknowledgments} 

J. P. would like to thank Lenny Susskind for the discussions that
initiated this work.
This work is supported by NSF grants PHY91-16964 and
PHY94-07194.

\end{document}